\documentclass[12pt]{article}
\pdfoutput=1
\usepackage{amsmath,amssymb,amsfonts,bbding,cancel}
\usepackage{graphicx,xcolor}

\newcommand{\be}{\begin{equation}} 
\newcommand{\ee}{\end{equation}} 
\newcommand{\bea}{\begin{eqnarray}}  
\newcommand{\eea}{\end{eqnarray}}

\newcommand{\SM}{\mathrm{SM}}

\begin{document}

%----------------------------------- TITLE AND AUTHORS -----------------------------------------%

\begin{flushright}
UG-FT-319/16 \\
CAFPE-189/16 \\

\today
\end{flushright}
\vspace*{5mm}
\begin{center}

\renewcommand{\thefootnote}{\fnsymbol{footnote}}

{\Large {\bf One-loop effective lagrangians after matching
}} \\
\vspace*{1cm}
{\bf F.\ del Aguila$^a$}\footnote{E-mail: faguila@ugr.es},
{\bf Z.\ Kunszt$^b$}\footnote{E-mail: kunszt@itp.phys.ethz.ch}
and
{\bf J.\ Santiago$^a$}\footnote{E-mail: jsantiago@ugr.es}

\vspace{0.5cm}

$^a$ Departamento de F\'{\i}sica Te\'orica y del Cosmos and CAFPE,\\
Universidad de Granada, E-18071 Granada, Spain \\
$^b$ Institute for Theoretical Physics, ETH Z\"urich, CH-8093 Z\"urich, Switzerland

\end{center}
\vspace{.5cm}

%--------------------------------------------- ABSTRACT ---------------------------------------------%
\begin{abstract}
We discuss the limitations of the covariant derivative expansion
prescription advocated 
to compute the one-loop Standard Model (SM) effective 
lagrangian when the heavy fields couple linearly to the SM. 
In particular, one-loop contributions resulting from the exchange of 
both heavy and light fields must be explicitly taken into account 
through matching because the proposed functional approach alone 
does not account for them. 
We review a simple case with a heavy scalar singlet of charge $-1$ 
to illustrate the argument. 
As two other examples where this matching is needed and this functional 
method gives a vanishing result, up to renormalization of the heavy
sector parameters, we re-evaluate the one-loop corrections 
to the T--parameter due to a heavy scalar triplet with vanishing 
hypercharge coupling to 
the Brout-Englert-Higgs boson and to a heavy vector-like quark singlet
of charged $2/3$  
mixing with the top quark, respectively. In all cases we make use of  
a new code for matching fundamental and 
effective theories in models with arbitrary heavy field additions.
\end{abstract}

\renewcommand{\thefootnote}{\arabic{footnote}}
\setcounter{footnote}{0}

%-------------------------------- DOCUMENT: INTRODUCTION ---------------------------------%

\section{Introduction}
\label{Intro}

The discovery of the Brout-Englert-Higgs (BEH) boson
\cite{Englert:1964et} at the LHC  
\cite{Aad:2012tfa} has completed the Standard Model (SM),
and with it the  
description of nature with a precision up to few per mille at the electro-weak 
scale \cite{delAguila:2011zs}. 
Moreover, the picture which seems to emerge from the stringent limits set 
by many of the LHC searches for new physics shows a gap 
up to the next layer of physics \cite{exotica} 
\footnote{\label{diphoton} Neglecting, for the time being, the diphoton excess observed 
at $\sim 750$ GeV by the LHC collaborations \cite{ATLASCMSdiphoton}.}. 
In this scenario one must use an effective lagrangian approach to 
study the low energy effects of possible heavy new resonances beyond the LHC 
reach: 
\begin{equation}
\mathcal{L}_{\mathrm{eff}}=\mathcal{L}_{\mathrm{SM}} + \sum_{n>4} \frac{1}{\Lambda^{n-4}}\mathcal{L}_n\ ,
\label{effectivelagrangian}
\end{equation}
where $\mathcal{L}_{\mathrm{SM}}$ is the SM lagrangian, $\Lambda$ the
next scale of new physics  
and $n$ the dimension of the local operators $\mathcal{O}^{(n)}_i$ entering in 
$\mathcal{L}_n = \sum_i \alpha^{(n)}_i \mathcal{O}^{(n)}_i$, with
$\alpha^{(n)}_i$ the corresponding  
Wilson coefficients.  
The classification of all operators $\mathcal{O}^{(6)}_i$ of dimension
6 in $\mathcal{L}_6$  
parameterizing the SM extensions in a model independent way was put forward  
some time ago \cite{Buchmuller:1985jz} 
\footnote{See also Ref. \cite{Weinberg:1979sa} for a discussion of the
  only dimension 5 operator built  
with SM fields,  
and which on the other hand also violates lepton number; and
Refs. \cite{Georgi:1991ch} for further developments on the  
dimension-6 lagrangian.}. 
The coefficients $\alpha^{(6)}_i$, which are expected to gather the
largest low-energy contributions  
of the heavy particles, do depend on the particular SM extension considered. 
As already noticed, the picture emerging from the LHC searches has
boosted the revival of the  
phenomenological interest in the theoretical prediction of the
coefficients of the effective lagrangian  
up to dimension 6 and up to one-loop order, to cope with the expected
experimental precision.  
With this purpose, the procedure to evaluate the contributions of new
(heavy) physics to this  
order has been revised in Ref. \cite{Henning:2014wua}
(see~\cite{Gaillard:1985uh} for related previous works), providing the
one-loop corrections for  
any SM addition with no \textit{linear} couplings to the SM (light) fields. 
In this work the evaluation of the one-loop contribution of a generic
heavy sector  
is reduced to an algebraic problem, getting rid of the difficulties
associated to  
the handling of the loop integrals. This is achieved by the clever
use of functional methods using the so called covariant derivative
expansion (CDE).
Its results readily apply to supersymmetric models with R--parity
\cite{Fan:2014axa},  
to models in which the heavy sector does not mix linearly with the
SM~\cite{Huo:2015exa} 
and, in general, to models with a (discrete) symmetry forbidding such linear terms \cite{Appelquist:2000nn,Cheng:2003ju}. 
This work has been also generalized to   
extend its range of applicability to the case of 
non-degenerate heavy field masses \cite{Drozd:2015rsp}. 
However, as already emphasized, although it has been claimed that the
method applies in general,   
it does not fully account for all quantum corrections when the SM addition involves heavy fields 
coupling linearly to the light (SM) fields. 
Since in this case there are one-loop corrections resulting 
from the exchange of both heavy and light fields within the loops which are not included in the 
algebraic result, which only accounts for the one-loop diagrams exchanging heavy particles alone 
\footnote{There can be one-loop contributions proportional to the
  linear couplings due to the  running of heavy particles alone, which
  are fully accounted for in the CDE method, see below.}. 
These contributions can be taken care of, however, performing a full matching 
with the proper local operators, as argued time ago in 
Ref. \cite{Witten:1975bh}.~\footnote{Such a matching could be in
  principle calculated using functional methods, as proposed, for
  example, in Refs.~\cite{Leon:1988za} and references there in.
  A generalization of the CDE prescription with this purpose is
  currently under investigation.\label{functional_matching}}

Let us be more precise about why this further matching is needed to recover the physical predictions of
the original  
theory. The straightforward application of the CDE
results in a different theory in the presence of a heavy 
sector coupling linearly to the SM. 
Indeed, the computation of the one-loop effective action ${\cal S}_{\mathrm{eff}}$ 
for the light (SM) fields $l$ by integrating out a heavy field $h$, 
\begin{equation}
e^{\ i {\cal S}_{\mathrm{eff}}[l]} = \int\ {\cal D} h\ e^{\ i {\cal S}[h,l]}\ ,
\label{effectiveaction}
\end{equation}
using the saddle point approximation requires solving the stationary 
equation for the action ${\cal S}$,  
\begin{equation}
\left. \frac{\delta {\cal S} [h,l]}{\delta h}\ \right|_{h=h_c} = 0\ ,
\label{stationaryequation}
\end{equation}
defining the heavy field classical solution $h_c$. 
Since, it is around this solution that the quantum fluctuations 
$H = h - h_c$ 
only enter {\it quadratically} in the path integral: 
\begin{align}
e^{\ i {\cal S}_{\mathrm{eff}}[l]} =& \int\ {\cal D} H\ e^{\ i {\cal S}[h_c+H,l]} 
\nonumber \\
=&\int \ {\cal D} H  e^{\ i \big( {\cal S}[h_c,l] + 
\frac{1}{2}\left. \frac{\delta^2 {\cal S} [h,l]}{\delta h^{
    2}}\ \right|_{h=h_c} H^2 +  
{\cal O} (H^3) \big)}\ .
\nonumber 
\label{actionexpansion}
\end{align}
However, in the presence of linear couplings of the heavy field to the
SM fields,   
${\cal L} [h,l] \supset h^\dagger J[l] + h.c.,$
the equation of motion for $h_c$ 
\begin{equation}
(D^2 + M^2 + U[l])\ h_c = J[l] + {\cal O} [h_c^2] \ ,
\end{equation}
where $D^2 = D_\mu D^\mu$ with $D_\mu$ the covariant derivative, $M$ is the $h$ 
mass and $U$ is the pertinent function of the light fields $l$, 
is solved by iteration (first equation below) 
also making use of an asymptotic expansion for the non-local operator 
${\cal O}^{-1} = - (D^2 + M^2 + U[l])^{-1}$ (second equation) 
\footnote{${\cal O}^{-1}$ has a finite radius of convergence and 
the series expansion is only valid for small values of the momenta and of the light fields.}
\begin{align}
h_c 
&\approx \frac{1}{D^2 + M^2 + U[l]} J[l] 
%\nonumber \\ &
= \frac{1}{M^2} \sum_{n=0}^\infty 
\left( - \frac{D^2 + U[l]}{M^2}\right)^n J[l] \ .
\label{classicalsolution}
\end{align}
But, this expansion is only applied for a series solution 
with a finite number of terms $N$, in which case the linear term is not 
eliminated but suppressed to the power $M^{-2N}$. 
In practice, one redefines 
\begin{equation}
h = H + \frac{1}{M^2} \sum_{n=0}^{N-1} 
\left( - \frac{D^2 + U[l]}{M^2}\right)^n J[l] \approx H + h_c \ , 
\label{fieldredefinition}
\end{equation}
which is a local, and then allowed, field redefinition. 
In such a case the linear coupling is only redefined (suppressed) to 
order $M^{-2N}$, 
\begin{equation}
{\cal L} [h,l] \supset - H^\dagger 
\left(\frac{D^2 + U[l]}{M^2}\right)^N J[l] + h.c.\ , 
\label{higherorderlinearterm}
\end{equation}
and cannot be ignored 
\footnote{In general, there can be further contributions to Eq. (\ref{higherorderlinearterm}) 
due to higher order terms in $h$ in the lagrangian 
but they result in higher order contributions in $M^{-1}$ at one loop.}. 
One may argue that in the limit $N \rightarrow \infty$ this coupling 
goes to zero, but then the expansion of ${\cal O}^{-1}$ is asymptotic 
and the resulting theory and physical predictions of both limits 
(integrating to arbitrary momenta for $N$ finite and taking $N \rightarrow \infty$ 
afterwards, or $N \rightarrow \infty$ and integrating to arbitrary momenta) 
are different. 
In summary, one can use Eq. (\ref{fieldredefinition}) keeping track 
of the linear term suppressed to the corresponding order, or use 
the quadratic contributions obtained by the CDE with the subsequent 
matching as indicated in Ref. \cite{Witten:1975bh}. 
We must insist again at this point when using the former approach that although the linear coupling is removed up to order $M^{-2N}$, it contributes to order $M^{-2}$ at one loop, 
as we will explicitly show in the example below. 

In the following section we work out an explicit example, reviewing a simple 
SM extension studied in full detail in Ref. \cite{Bilenky:1993bt}, the addition 
to the SM of one extra heavy charged scalar singlet $h$ 
of mass $M\ ( = \Lambda )$. 
We want to elaborate on the fact that the purely functional methods 
used in the CDE to compute the one-loop effective lagrangian 
(see Ref. \cite{Henning:2014wua})  
require further matching, as pointed out in
Refs. \cite{Witten:1975bh,Bilenky:1993bt}.  
The one-loop effective action computed with the proposed 
functional method is entirely governed by the terms in the full
lagrangian which are quadratic in the heavy fields. Furthermore, light
fields are kept constant through the calculation. Such
contributions correspond, diagrammatically, to one-loop diagrams in
which only heavy particles circulate in the loop. In contrast, the diagrammatic
calculation of the one-loop effective lagrangian by matching the fundamental 
and effective theories includes those contributions plus those in
which both heavy and light particles circulate in the loop (these
diagrams depend on the linear couplings of the heavy fields to the SM). 
Hence, these latter 
contributions do have to be taken into account but the CDE with its
present formulation 
does not incorporate them. (See footnote~\ref{functional_matching}.) 

As another example of physical interest where further matching is required 
after using the CDE recipe, we discuss in Section 3 the proper one-loop matching for the T--parameter in two other SM extensions with a heavy sector coupling linearly to the SM. 
In one case the extended model has an extra heavy scalar triplet coupling linearly to 
the BEH boson, and in the other one the SM is extended with 
one extra heavy vector-like quark singlet of charge 2/3 mixing with
the top quark.  
In both cases the CDE alone gives a vanishing one-loop
contribution (or gives a contribution that can be reabsorbed in the
renormalization of the mass of the heavy fields),  
in contrast with the straightforward diagrammatic computation. 
For this calculation we make use of \texttt{MatchMaker} \cite{matchmaker}, 
a new automated tool for evaluating tree-level and one-loop matching
conditions for arbitrary UV completions into effective
lagrangians.
The result for the examples worked out below agrees with that obtained in 
Ref. \cite{Khandker:2012zu} for the SM unbroken phase in the scalar
triplet case,  
and with the result in Refs. \cite{Lavoura:1992np,Carena:2006bn} for
the vector-like 
quark singlet addition.  
Section 4 is devoted to a summary. 
Technical details on the comparison with previous results in the
literature are relegated to an appendix.

\section{Extending the SM with a heavy charged scalar singlet}
\label{Example}

Let us assume the existence of a heavy scalar singlet $h$ of hypercharge $-1$ 
and of mass $M$, much larger that the electro-weak scale, as in Ref. \cite{Bilenky:1993bt}. 
Following it, we review in this section the discussion on the need of further matching of 
the effective field theory (EFT) obtained by the CDE integration of the heavy field with the 
fundamental theory to one loop, if both must describe the same physics at this order. 
However, it is not necessary in our case to go through the complete calculation of the one-loop 
effective lagrangian, which is already worked out in detail in Ref. \cite{Bilenky:1993bt}, 
but it is enough to show that the CDE does not 
account for a definite physical contribution at this order 
and hence, that it must be added through matching. 

The model we are interested in is described by the
lagrangian~\footnote{We have not explicitly written an allowed quartic
  term in 
  the heavy field, $\alpha |h|^4$, which plays no relevant r\^ole in
  our discussion.}   
\begin{equation}
\mathcal{L}=\mathcal{L}_{\mathrm{SM}}+ \mathcal{L}_h
=\mathcal{L}_{\mathrm{SM}}+ h^\dagger \mathcal{O} h +
h^\dagger J + J^\dagger h\ ,  
\label{hlagrangian}
\end{equation}
with $\mathcal{O}\equiv -D^2-M^2-\beta |\phi|^2$,
and $J\equiv \bar{\tilde{\ell}}_a f_{ab} \ell_b$.
$\phi$ is the SM scalar doublet and $\ell_a$ the SM lepton doublet of
flavor $a$.  
Besides, since $\tilde{\ell} = i \tau_2 \ell^c$, $f_{ab}$ is
antisymmetric in the flavor indices. 

In order to substantiate our point with this example we will identify 
first a physical amplitude for which the predictions in the 
fundamental theory and in the EFT obtained applying the CDE prescription are 
different. This means that the EFT mimicking the 
fundamental theory must be completed with the required local operators 
as shown in Ref. \cite{Bilenky:1993bt}. 
We will then show by analogy that this is needed because the 
implicit field redefinition used in this case when decomposing 
the heavy field as its classical counterpart plus its fluctuation corresponds 
to a non-allowed transformation, for the classical field definition involves 
a non-local operator which brings it to a different theory. 
This is made apparent observing that successive 
heavy field redefinitions in the fundamental theory with local transformations 
suppressing the linear coupling of the heavy field to SM fields up to order $M^{-2N}$ 
give the same physical results till the limit $N \rightarrow \infty$ is taken. 
Then, no linear heavy field coupling to light fields is present at all and 
the heavy field redefinition involves an infinite sum of terms expanding 
the non-local operator in Eq. (\ref{classicalsolution}), 
then requiring further matching.  
 
\subsection{Matching the EFT to the fundamental theory}

As worked out in Ref. \cite{Bilenky:1993bt}, the one-loop quantum corrections 
involving the $\beta$ parameter in Eq. (\ref{hlagrangian}) generate 
the one-loop ($1l$) effective lagrangian (at the renormalization scale
$\mu = M$ and  
omitting flavor indices)   
\begin{align}
\mathcal{L}_\beta^{(1l)}=\frac{1}{16\pi^2} 
&\Big\{
M^2 \beta (1+\Delta) |\phi|^2 + \frac{\beta^2}{2}
\Delta |\phi|^4
\nonumber \\
+ \frac{1}{M^2} \Big[ 
-&
\frac{\beta^3}{6} |\phi|^6 
  +\frac{\beta^2}{2} (\partial_\mu |\phi|^2)^2 
+ \frac{g^{\prime\ 2}\beta}{12} |\phi|^2 B_{\mu\nu}B^{\mu\nu} \Big]
\nonumber \\
&+\frac{\beta}{M^2} \Big[|\phi|^2 (\bar{\ell} f^\dagger f \cancel{D} \ell) +
  \mathrm{h.c.}\Big]\Big\}\ ,  
\label{L1beta}
\end{align}
where dimensional regularization with $d = 4 - 2\epsilon$ is used and 
$\Delta = 1/\epsilon-\gamma_E + \ln 4 \pi$.
$g^\prime$ and $B_{\mu\nu}$ stand for the hypercharge coupling and field strength, 
respectively. 
The first line in Eq. (\ref{L1beta}) renormalizes the SM lagrangian while 
the last two are part of the dimension 6 effective lagrangian. 
What matters to us, however, is that all the terms but the last one 
correspond to diagrams with only $h$ running in the loop, whereas the
last operator corresponds to a diagram with both heavy ($h$) and light
($\ell$) particles  
running in the loop. 
Hence, this latter contribution, which is also proportional to the
linear $h$ coupling  
in the full lagrangian in Eq. (\ref{hlagrangian}) (proportional to
$f$), is missing in the functional  
formalism. Therefore, it has to be computed through 
matching with the fundamental theory. In the effective theory there is no such one-loop contribution 
to the dimension--6 operator $|\phi|^2 \bar{\ell} \cancel{D} \ell$. 
We will then focus on this term.  

The matching can be performed by computing the relevant contribution
to the $\phi \phi^\dagger \to \ell \bar{\ell}$ amplitude, that we
denote
$\mathrm{i}\mathcal{M} = \bar{u}_a(p_2) \gamma_\mu u_b(p_1)
\mathrm{i}\mathcal{M}^\mu_{ab}$.
\begin{figure}
% Use the relevant command to insert your figure file.
% For example, with the graphicx package use
\begin{center}
  \includegraphics[width=0.5\columnwidth]{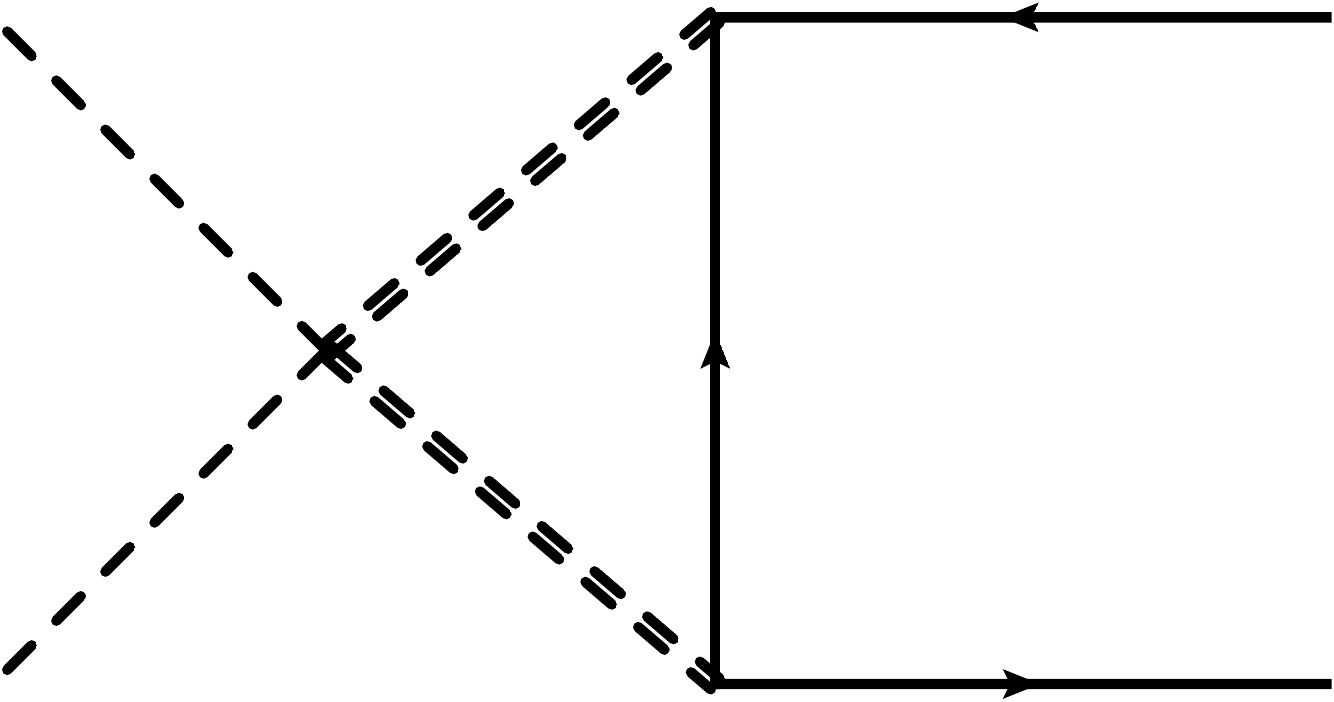}
% figure caption is below the figure
\caption{Feynman diagram contributing to
  $\mathrm{i}\mathcal{M}^\mu_{ab}$ in the full theory, see 
Eq. (\ref{iMmu:full:part1}).}
\label{fig:full:part1}       % Give a unique label
\end{center}
\end{figure}
The result in the fundamental theory reads
\begin{equation}
%\raisebox{-7mm}
%{\includegraphics[width=30mm]{arcadifull}}
%\sim 
\mathrm{i}\mathcal{M}_{ab}^\mu 
=-\frac{\mathrm{i}(d-4)(d-2)}{16\pi^2 d}\frac{\beta (f^\dagger f)_{ab}}{M^4} (p_1^\mu+p_2^\mu)
A(M^2) + \dots\ , 
\label{iMmu:full:part1}
\end{equation}
where the relevant Feynman diagram is shown in
Fig.~\ref{fig:full:part1}, the dots denote terms
proportional to higher powers of the external momenta. 
We define the tadpole integral  
\begin{align}
\frac{\mathrm{i}}{16\pi^2}
A(M^2) 
\equiv &
\mu^{2\epsilon} \int \frac{d^d k}{(2\pi)^d}
\frac{1}{k^2-M^2+\mathrm{i}\delta} 
\nonumber \\
=& \frac{\mathrm{i}}{16\pi^2} M^2 [\Delta +1 -
  \ln\left(\frac{M^2}{\mu^2} \right) + \mathcal{O}(\epsilon)]\ ,
\end{align}
where the $\Delta$ term is removed by counterterms in the $\overline{MS}$
scheme. The $d-4$ factor in Eq. (\ref{iMmu:full:part1}) guarantees a finite result 
that reads 
\begin{equation}
\mathcal{M}_{ab}^\mu = \frac{\beta(f^\dagger f)_{ab}}{16\pi^2 M^2} (p_1^\mu+p_2^\mu)
+ \dots\ .
\end{equation}
There is not such a contribution in the EFT obtained using functional methods 
and the beyond of the SM tree-level ($0l$) effective lagrangian 
\begin{equation}
\mathcal{L}^{(0l)}_{BSM} = -J^\dagger \mathcal{O}^{-1} J = \frac{J^\dagger J}{M^2} +
\mathcal{O}(M^{-4})\ .
\end{equation}
So both theories are not the same, unless we correct the latter with this 
additional matching.

Let us now discuss by analogy what happens when we redefine the heavy field  
in the fundamental theory by successive shifts, Eq. (\ref{fieldredefinition}), 
corresponding to keeping only a finite number of terms $N$ in the expansion of ${\cal O}^{-1}$ in 
Eq. (\ref{classicalsolution}).  
The transformation has unit Jacobian but involves a non-local operator in the limit $N
\rightarrow \infty$. 

\subsection{Heavy field redefinition at leading order}

Let us assume $N = 1$ in Eq. (\ref{fieldredefinition}). Then, the heavy field $h$, 
named $H$ after redefining it, equals to first order 
\begin{equation}
h=H+\frac{J}{M^2} 
\end{equation}
and hence, the heavy lagrangian in Eq. (\ref{hlagrangian}) reads 
\begin{equation}
\mathcal{L}_H^{{\rm LO}} = H^\dagger \mathcal{O} H + \frac{J^\dagger J}{M^2}
+ \frac{1}{M^2}\left( H^\dagger \hat{\mathcal{O}} J +
  \mathrm{h.c.}\right)
+\frac{J^\dagger \hat{\mathcal{O}} J}{M^4}\ ,
\end{equation}
with
\begin{equation}
\hat{\mathcal{O}}\equiv \mathcal{O} + M^2 = -D^2 - \beta |\phi|^2\ .
\end{equation}
As required, the linear coupling is now suppressed up to order 
$M^{-2}$. However, this linear coupling, despite its 
higher-order suppression, still provides the \textit{same} physical 
amplitudes, of order $M^{-2}$, because we have only 
performed an allowed (local) field redefinition. Indeed, 
focusing again on the amplitude $\phi \phi^\dagger \to 
\ell \bar{\ell}$,
the relevant (new) Feynman rules read now 
\begin{equation}
\raisebox{-7.5mm}
{\includegraphics[width=15mm]{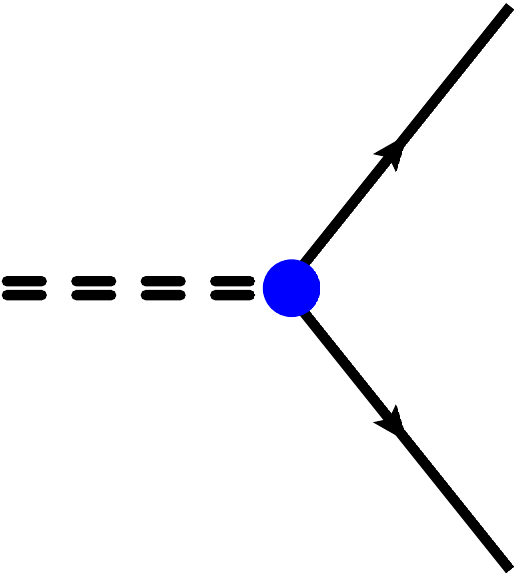}}
=\mathrm{i}\frac{p_H^2}{M^2}f^\dagger_{ab}\ ,
\qquad
\raisebox{-7.5mm}
{\includegraphics[width=15mm]{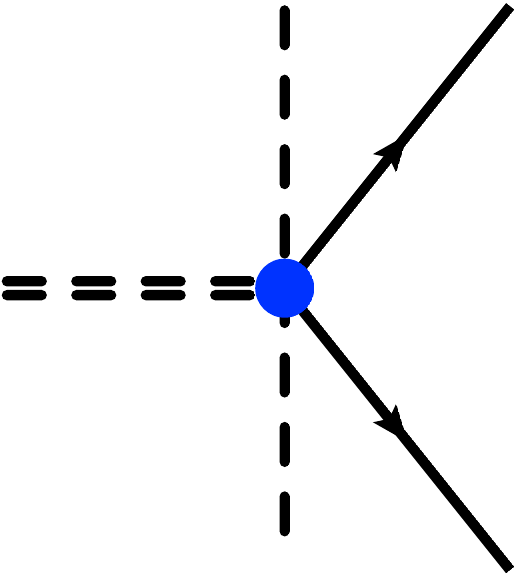}}
=-\mathrm{i}\frac{\beta}{M^2}f^\dagger_{ab}\ ,
\end{equation}
where the blue dot stands for an order $M^{-2}$ coupling and $p_H$ is
the $H$ momentum.
\begin{figure*}
% Use the relevant command to insert your figure file.
% For example, with the graphicx package use
\begin{center}
  \raisebox{7mm}{\includegraphics[width=0.3\columnwidth]{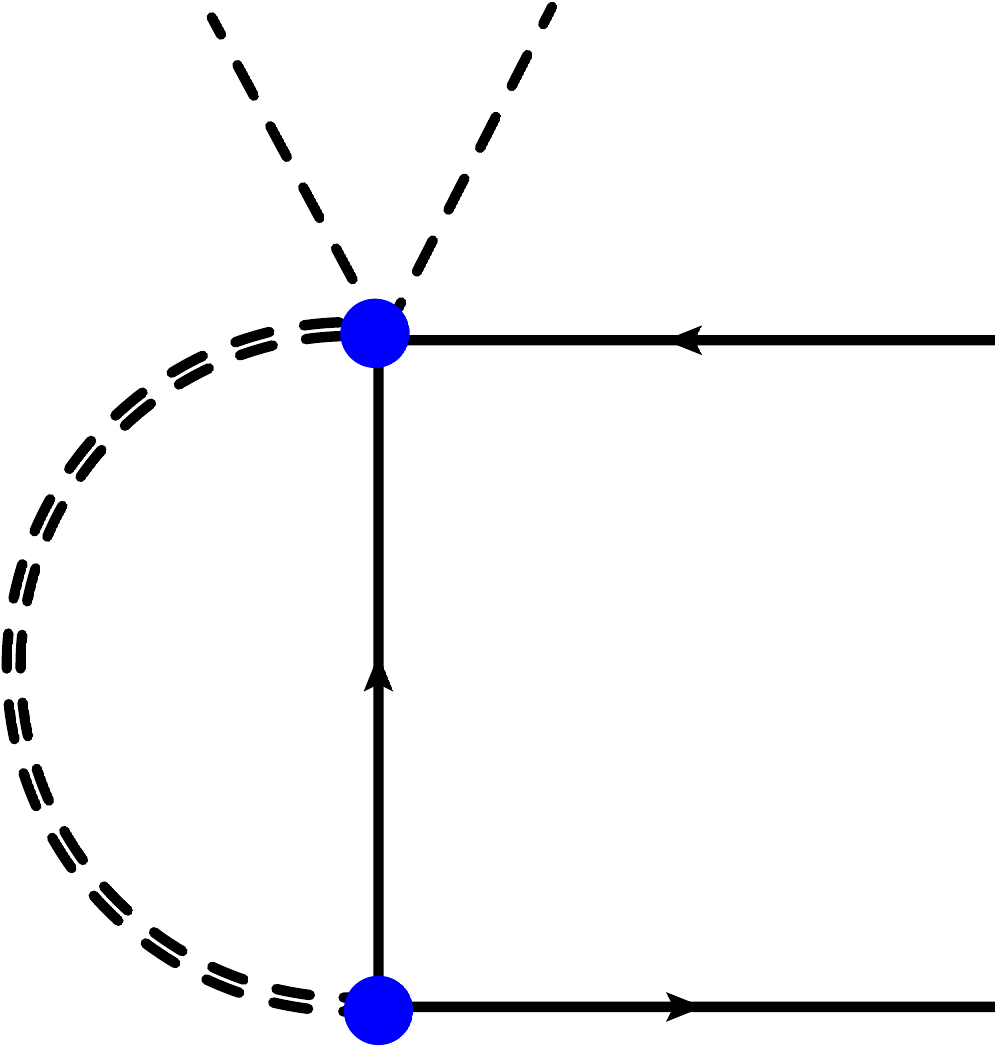}}
\hfil  
\includegraphics[width=0.3\columnwidth]{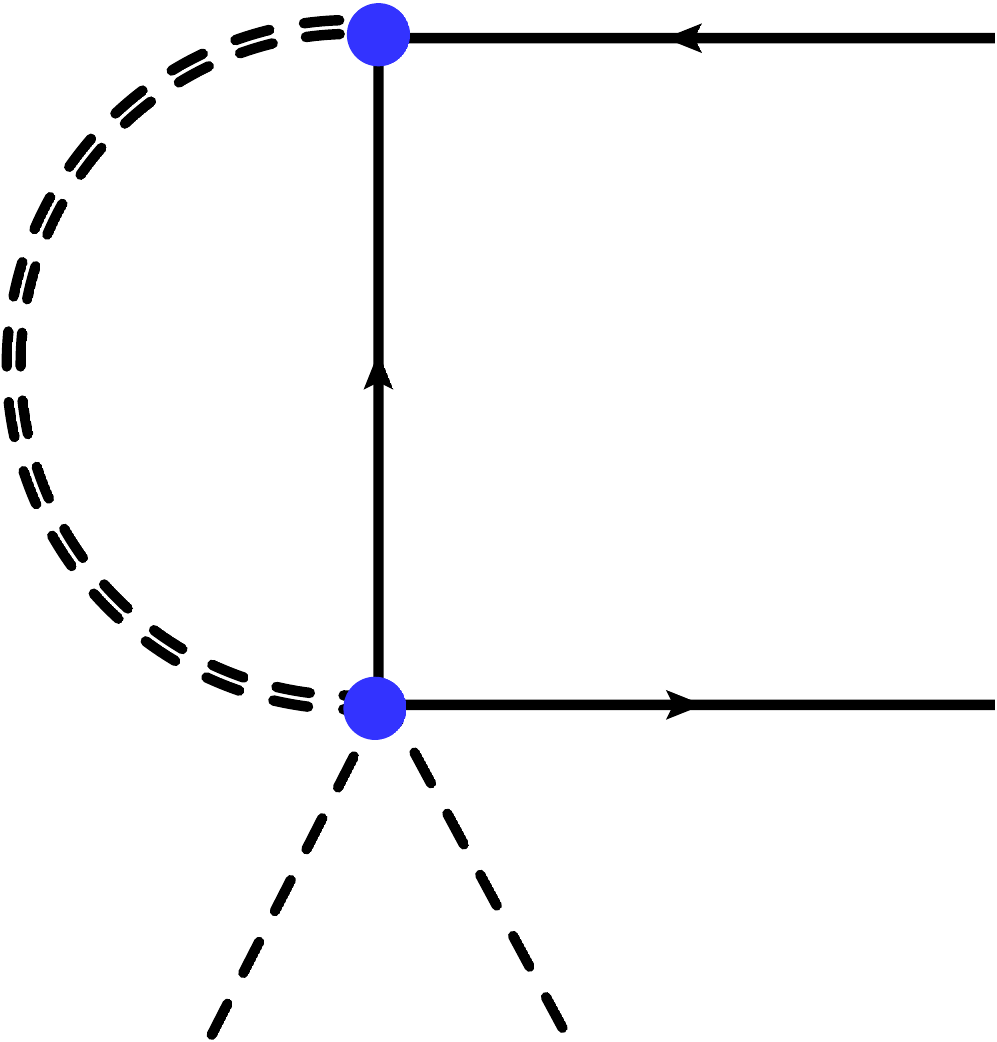}
\hfil  
\raisebox{8mm}{\includegraphics[width=0.38\columnwidth]{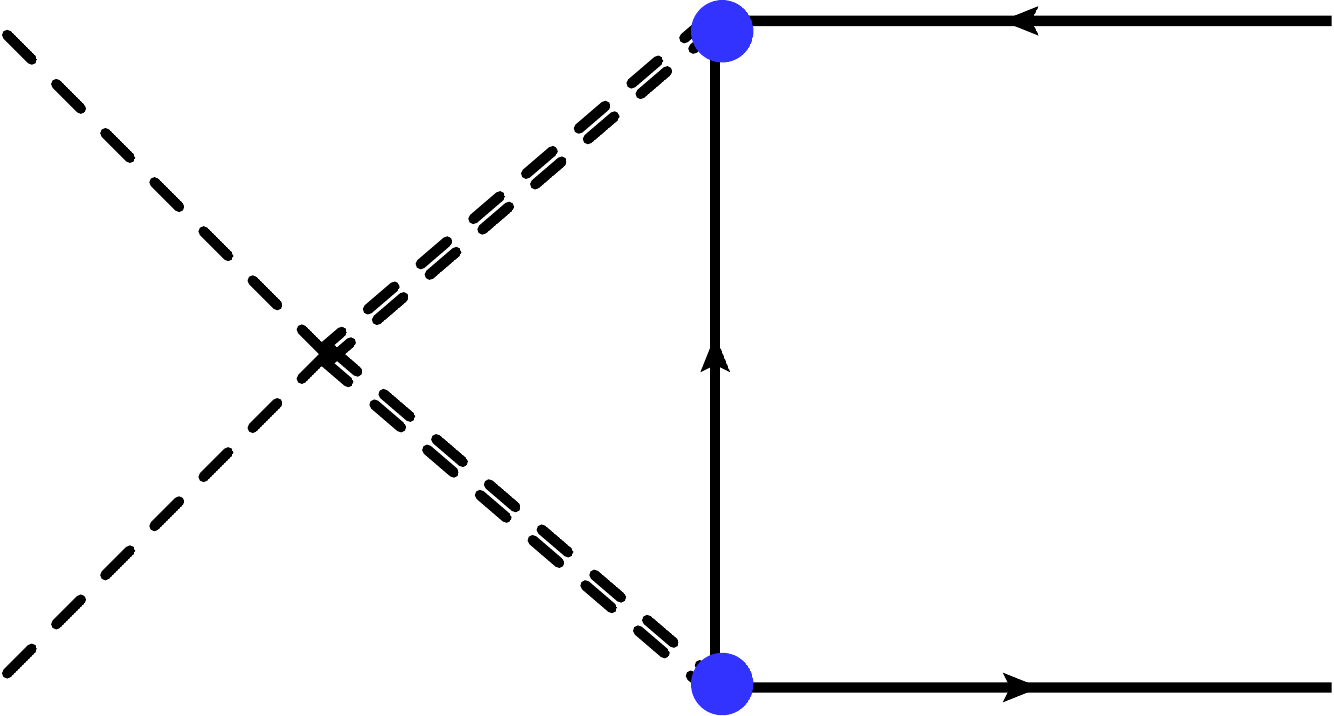}}
% figure caption is below the figure
\caption{Feynman diagrams contributing to
  $\mathrm{i}\mathcal{M}^\mu_{1}$ (left),
  $\mathrm{i}\mathcal{M}^\mu_{2}$ (center) and
  $\mathrm{i}\mathcal{M}^\mu_{3}$ (right)
 in 
Eq. (\ref{iMmu3:eff:LO:part1})
(and Eq. (\ref{iMmu3:eff:NLO:part1}) but with the blue dots replaced
with red squares).}
\label{fig:eff:LO:part1}       % Give a unique label
\end{center}
\end{figure*}
The three diagrams that contribute to this amplitude are shown in
Fig. \ref{fig:eff:LO:part1} (we omit flavor indices) with the
following result
\begin{align}
%\raisebox{-7mm}
%{\includegraphics[width=20mm]{iMmu1effLO}}
%&\sim& 
\mathrm{i}\mathcal{M}^{{\rm LO}\,\mu}_1 
&=\frac{\mathrm{i}}{16\pi^2}
\frac{4(d-2)}{d}\frac{\beta (f^\dagger f)}{M^4} p_1^\mu
A(M^2) + \dots\ , 
\nonumber \\
%\raisebox{-12mm}
%{\includegraphics[width=20mm]{iMmu2effLO}}
%&\sim& 
\mathrm{i}\mathcal{M}^{{\rm LO}\,\mu}_2 
&=\frac{\mathrm{i}}{16\pi^2}
\frac{4(d-2)}{d}\frac{\beta (f^\dagger f)}{M^4} p_2^\mu
A(M^2) + \dots\ , \nonumber 
\\
%\raisebox{-5mm}
%{\includegraphics[width=20mm]{iMmu3effLO}}
%&\sim& 
\mathrm{i}\mathcal{M}^{{\rm LO}\,\mu}_3 
&=-\frac{\mathrm{i}}{16\pi^2}
(d-2)\frac{\beta (f^\dagger f)}{M^4} (p_1^\mu+p_2^\mu)
A(M^2) + \dots\ . \label{iMmu3:eff:LO:part1}
\end{align}
Although each of the three amplitudes is separately divergent, the sum is finite and
equals, as expected, Eq. (\ref{iMmu:full:part1}): 
\begin{equation}
\mathcal{M}^\mu = \mathcal{M}^{{\rm LO}\,\mu}_1 + \mathcal{M}^{{\rm LO}\,\mu}_2 +
\mathcal{M}^{{\rm LO}\,\mu}_3\ .
\end{equation}

\subsection{Heavy field redefinition at next to leading order}

Let us repeat the exercise to next order in $M^{-2}$. 
The heavy field redefinition reads now 
\begin{equation}
h=H+\frac{J}{M^2}+\frac{\hat{\mathcal{O}}}{M^4} J\ ,
\end{equation}
which leaves the lagrangian 
\begin{align}
\mathcal{L}_H^{{\rm NLO}} 
&= H^\dagger \mathcal{O} H + \frac{J^\dagger J}{M^2}
- \frac{1}{M^4}\left( H^\dagger \hat{\mathcal{O}}^2 J +
  \mathrm{h.c.}\right)
\nonumber \\
&+\frac{J^\dagger \hat{\mathcal{O}} J}{M^4}
+\frac{J^\dagger \hat{\mathcal{O}}^2 J}{M^6}
+\frac{J^\dagger \hat{\mathcal{O}}^3 J}{M^8}\ .
\end{align}
The linear coupling is now suppressed up to order $M^{-4}$. 
The relevant (new) Feynman rules read at this order 
\begin{align}
\raisebox{-7.5mm}
{\includegraphics[width=15mm]{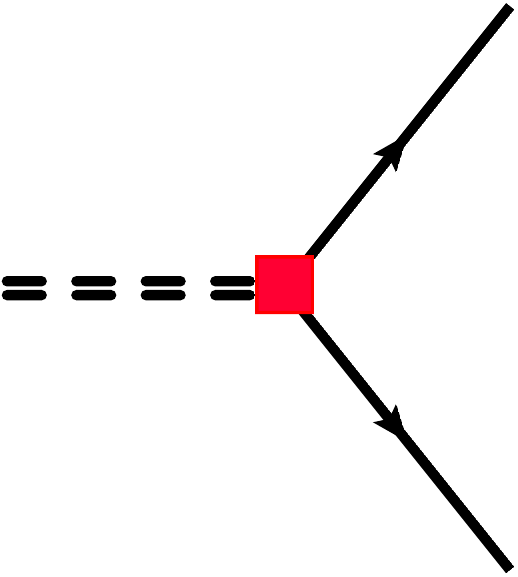}}
&
=-\mathrm{i}\frac{p_H^4}{M^4}f^\dagger_{ab}\ ,
\nonumber \\
\qquad
\raisebox{-7.5mm}
{\includegraphics[width=15mm]{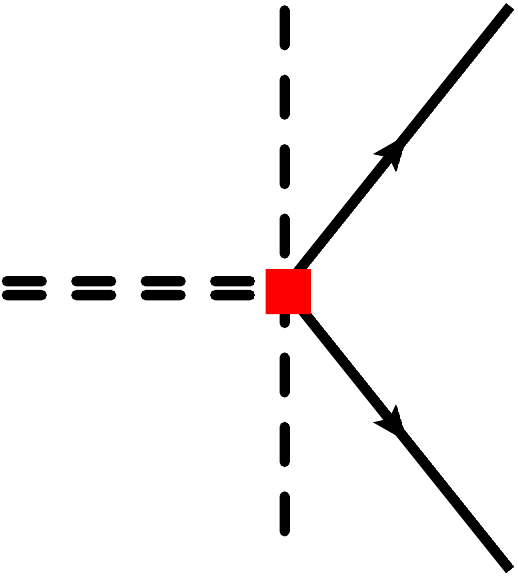}}
&
=\mathrm{i}\frac{\beta}{M^4}[p_H^2+(p_{\ell_1}-p_{\ell_2})^2]f^\dagger_{ab}\ ,
\end{align}
where the red square denotes a coupling of order $M^{-4}$ and 
the fermion momenta, $p_{\ell_1,\ell_2}$ follow the particle flow. 
%\begin{figure*}
% Use the relevant command to insert your figure file.
% For example, with the graphicx package use
%\begin{center}
%  \raisebox{7mm}{\includegraphics[width=0.3\columnwidth]{iMmu1effNLO}}
%\hfil  
%\includegraphics[width=0.3\columnwidth]{iMmu2effNLO}
%\hfil  
%\raisebox{8mm}{\includegraphics[width=0.38\columnwidth]{iMmu3effNLO}}
%% figure caption is below the figure
%\caption{Feynman diagrams contributing to
%  $\mathrm{i}\mathcal{M}^\mu_{1}$ (left),
%  $\mathrm{i}\mathcal{M}^\mu_{2}$ (center) and
%  $\mathrm{i}\mathcal{M}^\mu_{3}$ (right)
% in 
%Eq. (\ref{iMmu3:eff:NLO:part1}).}
%\label{fig:eff:NLO:part1}       % Give a unique label
%\end{center}
%\end{figure*}
The same three diagrams contribute to the
$\phi \phi^\dagger \to \ell \bar{\ell}$  
amplitude but with different couplings and weights
(Fig.~\ref{fig:eff:LO:part1} but with the blue dots replaced with red 
squares):  
\begin{align}
% \raisebox{-7mm}
%{\includegraphics[width=20mm]{iMmu1effNLO}}
%&\sim& 
\mathrm{i}\mathcal{M}^{{\rm NLO}\,\mu}_1
&=\frac{8\mathrm{i}}{16\pi^2}
\frac{(d-1)p_1^\mu-p_2^\mu}{d}\frac{\beta (f^\dagger f)}{M^4} 
A(M^2) + \dots\ , \nonumber 
\\
 %\raisebox{-12mm}
%{\includegraphics[width=20mm]{iMmu2effNLO}}
% &\sim& 
\mathrm{i}\mathcal{M}^{{\rm NLO}\,\mu}_2
&=\frac{8\mathrm{i}}{16\pi^2}
\frac{(d-1)p_2^\mu-p_1^\mu}{d}\frac{\beta (f^\dagger f)}{M^4} 
A(M^2) + \dots\ , \nonumber 
\\
%\raisebox{-5mm}
%{\includegraphics[width=20mm]{iMmu3effNLO}}
% &\sim& 
\mathrm{i}\mathcal{M}^{{\rm NLO}\,\mu}_3
&=%\frac{\mathrm{i}}{16\pi^2}
-\frac{\mathrm{i}(d+4)(d-2)}{16\pi^2 d}\frac{\beta (f^\dagger f)}{M^4} (p_1^\mu+p_2^\mu)
A(M^2) + \dots\ . \label{iMmu3:eff:NLO:part1}
\end{align}
We find again that the three contributions are separately divergent
but their sum is finite and exactly agrees with Eq. (\ref{iMmu:full:part1}), 
\begin{equation}
\mathcal{M}^\mu = \mathcal{M}^{{\rm NLO}\,\mu}_1 + \mathcal{M}^{{\rm NLO}\,\mu}_2 +
\mathcal{M}^{{\rm NLO}\,\mu}_3\ .
\end{equation}
As both calculations in the last two subsections show, we recover the 
physical one-loop amplitude $\phi \phi^\dagger \to 
\ell \bar{\ell}$ to
whatever order 
$M^{-2N}$ we suppress the linear coupling of the heavy field to the SM
as long as the heavy field 
redefinition is allowed (local), \textit{i.e.} $N < \infty$. 
But if the $N \rightarrow \infty$ limit is formally 
taken at the lagrangian level, there is no linear coupling of the heavy field left at all, and 
we have to deal with a non-local operator (transformation) and a different theory 
with different physical predictions.

\section{SM extensions with heavy scalars and fermions}
\label{Extensions}

The issue raised in the previous section also applies to any SM
extension with heavy fields coupling linearly to the light fields. 
In the following we provide the tree-level and one-loop matching
conditions 
relevant for the calculation of the T--parameter \cite{Peskin:1991sw} in two
of these SM extensions. 
In both cases the one-loop contribution to the T--parameter entirely arises 
from terms linear in the heavy fields. Hence, these contributions are
missing in the one-loop effective  
lagrangian obtained by functional methods only.~\footnote{There is a
  contribution in our first example proportional to the linear
  couplings that arise from loops involving only heavy particles and 
  therefore, it is correctly accounted for in the
  CDE~\cite{Henning:2014wua}. This term can be
  reabsorbed by a renormalization of the heavy particle mass as we
  discuss below.}

Using the language of the SM effective lagrangian, the T--parameter,
defined as the correction to the SM contribution and 
absorbing the electro-magnetic coupling constant $\alpha_{EM} = e^2 / 4 \pi$ 
\cite{Barbieri:2004qk}, 
 can
be written from the Wilson coefficient $\alpha_1$ of the dimension-6
effective operator 
(we omit the superscript indicating the operator dimension in the following)
\begin{equation}
\mathcal{O}_1 = \left| \phi^\dagger D_\mu \phi \right|^2\quad \mbox{ as } 
\quad \Delta {\hat {\rm T}} = -\alpha_{1} v^2, 
\label{T:from:alphaO1}
\end{equation}
with $v=174$ GeV the SM vacuum expectation value.

We use \texttt{MatchMaker} \cite{matchmaker}, 
an automated tool that performs tree-level and one-loop matching for
arbitrary extensions of the SM, for the actual matching. 
This is performed off-shell,
which means that all independent (including redundant) operators with
four Higgs bosons and two covariant derivatives have to be
considered. In particular, we use the basis 
\begin{equation}
\mathcal{O}_2=\phi^\dagger \phi\ \partial^2(\phi^\dagger \phi)\ ,
\quad
\mathcal{R} = \phi^\dagger \phi\ \phi^\dagger D^2 \phi\ , 
\end{equation}
where the hermitian conjugate of $\mathcal{R}$ has to be also included in
the effective lagrangian, 
with $\alpha_{\mathcal{R}^\dagger}= \alpha_{\mathcal{R}}^\ast$ .
The matching is performed by computing the one-light-particle-irreducible 
(1LPI) contributions to the Green's function 
\begin{equation}
\langle H_1 H_1^\ast H_2 H_2^\ast \rangle \ , \quad {\rm with}\ 
\phi = \begin{pmatrix} H_1 \\ H_2 \end{pmatrix}\ , 
\end{equation}
in the full and effective theories.
In the effective theory this amplitude reads
\begin{align}
\mathcal{M}=&
-
\alpha_{1} \big[p_2\cdot (p_3-p_4)-p_3\cdot(p_3+p_4)\big]
\nonumber \\ &
+2 \alpha_{2} (p_3+p_4)^2 
+ \alpha_{\mathcal{R}}^\ast \big[p_2^2+p_4^2 \big]
\nonumber \\ &
+ \alpha_{\mathcal{R}}
\big[(p_2+p_3)^2+(p_3+p_4)^2 + 2 p_2 \cdot p_4 \big],
\label{iM:H1H1barH2H2bar}
\end{align} 
where all momenta are considered incoming and we have already used
momentum conservation to eliminate $p_1$. The corresponding
calculation in the relevant extension of the SM will fix the matching conditions, 
up to possible wave function renormalization of the SM fields (see below). 

\subsection{SM extension with a heavy scalar triplet}
\label{Scalar triplet}

In this subsection we consider the first example discussed in
Ref. \cite{Henning:2014wua}. The same model was previously considered 
in \cite{Khandker:2012zu}, 
where the Wilson coefficients were computed by means of matching conditions. 
The SM addition consists of an extra real
scalar in the $(1,3,0)$ representation of the SM gauge symmetry group 
$SU(3)_C \times SU(2)_L \times U(1)_Y$. Denoting by 
$\Phi^a$ its three $SU(2)_L$ components, with $a = 1,2,3$, the heavy field 
lagrangian reads 
\begin{align}
\mathcal{L}_{\Phi}
&= \frac{1}{2} D_\mu \Phi^a D^\mu \Phi^a 
- \frac{1}{2} M^2 \Phi^a \Phi^a - \frac{\lambda_\Phi}{4} (\Phi^a \Phi^a)^2 
\nonumber \\
&
+ \kappa \phi^\dagger \sigma^a \phi \Phi^a - \eta \phi^\dagger \phi
\Phi^a \Phi^a\ , 
\end{align}
where $\phi$ is the SM scalar doublet with quartic coupling 
$-\lambda (\phi^\dagger \phi)^2$ 
and $\sigma^a$ are the Pauli matrices. 

The 1LPI contributions to the Green's function
$\langle H_1 H_1^\ast H_2 H_2^\ast \rangle$ 
reproduce the momentum structure of the effective theory
calculation in Eq. (\ref{iM:H1H1barH2H2bar}) with the tree-level 
Wilson coefficients 
\begin{equation}
\alpha^{(0l)}_{1}=-4\alpha^{(0l)}_{2} 
= 2\alpha^{(0l)}_{\mathcal{R}}
= -2 \frac{\kappa^2}{M^4}\ , 
\end{equation}
and the one-loop ones 
\begin{align}
\alpha_{1}^{(1l)}= &~\frac{\kappa^2}{16\pi^2 M^4}
\bigg( - \frac{\kappa^2}{M^2} +16 \eta -6 \lambda -20 \lambda_\Phi 
-\frac{5}{4}g_2^2\bigg), \nonumber \\
\alpha_{2}^{(1l)}=&~ \frac{1}{16\pi^2 M^2}
\bigg [ 
-\frac{\eta^2}{2} 
-\frac{13}{4} \frac{\kappa^4}{M^4}
\nonumber \\
&\qquad -\frac{\kappa^2}{M^2} \bigg(
\frac{7}{2} \eta
+\frac{15}{16} g_1^2
+\frac{25}{16} g_2^2
-6 \lambda -5 \lambda_\Phi \bigg) \bigg],\nonumber \\
\alpha_{\mathcal{R}}^{(1l)}= 
&~\frac{\kappa^2}{16\pi^2 M^4}
\bigg(\frac{21}{2} \eta 
+ \frac{5}{8} g_1^2 
+ \frac{5}{4} g_2^2 
-\frac{25}{2} \lambda 
-10\lambda_\Phi 
+\frac{21}{4} \frac{\kappa^2}{M^2}
\bigg)\ .
\end{align}

As mentioned above, the wave function renormalization of the SM fields 
must be also taken into account. In our case, the heavy triplet also contributes  
to the SM scalar doublet kinetic term at the loop level: 
\begin{equation}
  \mathcal{L}_{\SM}=\left(1 +\frac{3}{2} \frac{\kappa^2}{16 \pi^2 M^2}
 \right) |\partial_\mu \phi|^2 + \ldots\ ,
\end{equation}
This term can be reabsorbed by a $\phi$ redefinition
\begin{equation}
\phi \to \left ( 1-\frac{3}{4} \frac{\kappa^2}{16\pi^2 M^2} \right )
\phi 
\end{equation}
which, due to the non-vanishing tree-level contribution to the
corresponding effective operators, gives an extra contribution to
the one-loop Wilson coefficients. 
Focusing on $\mathcal{O}_1$, we get
\begin{equation}
\alpha_{1}^{(1l)} 
\to \alpha_{1}^{(1l)}  +\frac{6 \kappa^4}{16\pi^2 M^6}\ .
\end{equation}
Thus, our final result for $\alpha_{1}^{(1l)}$ reads
\begin{equation}
\alpha_{1}^{(1l)}= \frac{\kappa^2}{16\pi^2 M^4}
\left( 5 \frac{\kappa^2}{M^2} +16 \eta -6 \lambda -20 \lambda_\Phi 
-\frac{5}{4}g_2^2\right) .
\end{equation}
As it is apparent, it is proportional to the
linear coupling $\kappa$. In fact, except for the term proportional to
$\lambda_\Phi$, which arises solely from heavy particles running in
the loop and therefore appears in the CDE, all the remaining
ones are absent in the functional method calculation
alone~\cite{Henning:2014wua}. 
This is the result we were looking for. In order to find literal agreement 
with the calculation in \cite{Khandker:2012zu} by Khandker, Li and
Skiba (KLS), we have to remember that our quartic coupling for the BEH scalar 
doublet $\lambda = \lambda_{KLS}/4$ and that they use the
one-loop renormalized mass for the heavy triplet (as opposed to the
tree-level one which we are using). Their relation, which can
be found by computing the 1PI contribution to the $\Phi^a$ two-point
function, is 
\begin{equation}
M_{KLS}^2 = M^2\left ( 1
+\frac{2g_2^2-5\lambda_\Phi}{16 \pi^2}\right) ,
\end{equation}
which in turn gives the extra contribution from $\alpha_{1}^{(0l)}$ 
to its one-loop counterpart: 
\begin{equation}
\alpha_{1}^{(1l)} \to 
\alpha_{1}^{(1l)}  +\frac{(20\lambda_\Phi-8 g_2^2)\kappa^2}{16\pi^2
  M^4_{KLS}}\ .
\end{equation}
Adding all these contributions (and using the BEH quartic coupling normalization 
in \cite{Khandker:2012zu}), we obtain
\begin{align}
\alpha_{1\ KLS}^{(1l)} = \frac{\kappa^2}{16\pi^2 M_{KLS}^4}
\bigg(
& 
5 \frac{\kappa^2}{M_{KLS}^2} +16 \eta 
%\nonumber \\&
-\frac{3}{4} \lambda_{KLS}  
-\frac{37}{4}g_2^2\bigg) , 
\end{align}
which coincides with the result obtained in Ref. \cite{Khandker:2012zu}.

\subsection{SM extension with a heavy vector-like quark singlet}
\label{Fermion singlet}

Our last example is the extension of the SM with a vector-like quark 
$T$ in the $(3,1,2/3)$ representation of the SM gauge group. 
In this case the T--parameter is only generated at one-loop order, 
which was originally computed in \cite{Lavoura:1992np}
(see also \cite{Carena:2006bn,Anastasiou:2009rv} for extensions to
vector-like quarks  
in arbitrary representations). 
The lagrangian involving the heavy field reads
\begin{equation}
\mathcal{L}_{T} = \overline{T} (\mathrm{i} \cancel{D} -
M) T 
- \Big[ \lambda_T ~ \overline{q_L} 
\tilde{\phi} T_R + \mathrm{h.c.}\Big]\ ,
\end{equation}
with $T = T_L + T_R$, and $L$ and $R$ stand for left- and right-handed
fermions, respectively.  

The 1LPI calculation of the Green's function\\ 
$\langle H_1 H_1^\ast H_2 H_2^\ast \rangle$ 
in the full model has no tree-level contribution but the
one-loop values for the Wilson coefficients 
\begin{align}
\alpha_{1}^{(1l)}=&~ 
\frac{N_C |\lambda_T|^2}{16\pi^2 M^2}
\left( \frac{1}{2}\lambda_t^2 -\frac{1}{2} |\lambda_T|^2\right), 
\nonumber \\
\alpha_{2}^{(1l)}=& ~
\frac{N_C |\lambda_T|^2}{16\pi^2 M^2}
\left( \frac{3}{2}\lambda_t^2 -\frac{1}{3} |\lambda_T|^2\right), 
\nonumber \\
\alpha_{\mathcal{R}}^{(1l)}=&~ 
\frac{N_C |\lambda_T|^2}{16\pi^2 M^2}
\left( -\frac{1}{2}\lambda_t^2 +\frac{1}{2} |\lambda_T|^2\right), 
\label{alphaR1:T}
\end{align}
where $N_C=3$ for a quark and $\lambda_t$ is the corresponding 
top Yukawa coupling (we neglect all other SM Yukawa couplings). 

In this case, since the tree-level contribution vanishes, wave
function renormalization gives no further contributions at one
loop. Previous calculations of the T--parameter in this model have been
performed at the electroweak scale. In order to compare with our calculation, 
we have to run the Wilson coefficients down to the
top quark mass and integrate out the top quark with the anomalous
couplings induced by the heavy fermion. We present the details of 
this computation in Appendix \ref{appendix}, showing the agreement
with previous results.  

\section{Conclusions}
\label{Conclusions}

The LHC picture of nature seems to confirm a significant gap between the 
SM (light fields) and the new layer of physics (heavy fields). 
This makes the use of EFT compulsory in order to describe (bound) possible small 
deviations from the SM predictions in the high energy tail of the experimental 
distributions. 
Although an EFT with SM symmetries and light fields and arbitrary dimension--6 
operators built with them must be in general enough to describe such a scenario 
(neglecting in this context neutrino masses and the new physics associated to 
them), it is mandatory to recognize the relations among the different Wilson 
coefficients of these operators to identify the particular new physics realized in 
nature. 
With this purpose, different calculations of the one-loop contributions to the 
Wilson coefficients of the dimension--6 operators for different SM extensions have 
been made available using the CDE \cite{Henning:2014wua}. 
SM additions with linear couplings to light fields are treated in the same way as 
those without them, but in the former case the general results miss extra contributions which 
must be added by further matching with the specific fundamental theory \cite{Witten:1975bh}.  
Hence, although there are many phenomenologically relevant SM extensions 
without such linear terms, as supersymmetric theories with unbroken R-parity 
or models with universal extra dimensions, and in general theories with a 
discrete symmetry requiring interactions with only an even number of heavy fields, 
also many phenomenologically relevant theories include heavy fields with linear 
couplings to the SM, and they demand further treatment. 

This problem and its solution were pointed out some time ago \cite{Witten:1975bh}, and 
definite examples have been also worked out in detail \cite{Bilenky:1993bt}. 
The CDE does not include those linear couplings in loops, 
in contrast with the fundamental theory. What means that the corresponding 
contributions must be added through matching. (See
footnote~\ref{functional_matching}.) In this paper we elaborate on
this issue. Noticing first in a simple case with a heavy charged scalar singlet 
that the problem arises when we perform the non-local heavy field redefinition 
implicit in this functional treatment of theories with linear couplings of 
heavy fields to the SM. The fundamental theory (lagrangian) transformed by 
a local heavy field redefinition expressible as a series with a finite number 
of terms (local operators) $N$ in general gives the same physical predictions, 
till the infinite limit $N \rightarrow \infty$ is taken and the series becomes 
the asymptotic expansion of a non-local operator with a finite radius of 
convergence. 
Then, the physical predictions, as well as the resulting theory, are in general different, 
up to the proper matching. 

We have also discussed the beyond the SM contributions to the T--parameter in two other 
SM extensions with linear couplings of the heavy sector to the light fields, 
providing the missing pieces in the CDE.
They result from the addition of a heavy scalar triplet with vanishing
hypercharge and 
of a heavy vector-like quark of charge 2/3, respectively. 
As a matter of fact, the CDE gives in both cases a contribution that
is either vanishing or can be reabsorbed in the physical definition of
the heavy field mass. We obtain perfect agreement with previous calculations 
in both cases, with Ref. \cite{Khandker:2012zu} in the scalar triplet case and
with Ref. \cite{Lavoura:1992np} in the vector-like quark one. 
At any rate, all SM extensions with linear couplings between the 
heavy and light sectors can require such an extra matching, 
which can be in general of phenomenological interest (sizable), too.

For our explicit calculations we have made use of the new code
\texttt{MatchMaker}  
\cite{matchmaker}, aimed at automated calculation of tree-level and
one-loop matching conditions in arbitrary extensions of the SM.
Details of the code and its use will be presented elsewhere
\cite{matchmaker}.  

\section*{Acknowledgments}

We thank useful discussions with C. Anastasiou, 
J. R. Espinosa and A. Lazopoulos, and comments and a careful 
reading of the manuscript by M. P\'erez-Victoria and
A. Santamar{\'\i}a. We also thank B. Henning, X. Lu and H. Murayama
for useful correspondence regarding their work.   
This work has been supported
in part by the European 
Commission through the contract PITN-GA-2012-316704 (HIGGSTOOLS), by
the Ministry 
of Economy and Competitiveness (MINECO), under grant number 
FPA2013-47836-C3-1/2-P (fondos FEDER), 
and by the Junta de Andaluc{\'\i}a grants FQM
101 and FQM 6552. 

\appendix

\section{T--parameter at the electroweak scale}
\label{appendix}

In this appendix we show that our result for
$\alpha_{1}$ in the SM extension with an extra
vector-like quark singlet of hypercharge $2/3$, Eq. (\ref{alphaR1:T}),
agrees with previous calculations of the T--parameter in this model 
\cite{Lavoura:1992np}. Previous computations evaluate the T--parameter at the
electroweak scale directly in the physical basis 
(after electroweak symmetry breaking). 
The exact result, in the limit of large $M$ and only keeping up to 
$v^2/M^2$ terms reads \cite{Carena:2006bn} 
\begin{equation}
\Delta \hat{\rm T} = \frac{N_C}{32 \pi^2} \frac{v^2}{M^2}
\left[ |\lambda_T|^{4}+ 2\lambda_t^2 |\lambda_T|^{2} \left(\log
\frac{M^2}{m_t^2}-1\right)  \right], \label{DeltaThat_from_full}
\end{equation} 
where $m_t$ is the top mass.

In order to reproduce this result in our effective theory approach,
using Eq. (\ref{T:from:alphaO1}), we need to compute the corresponding
Wilson coefficient at the electroweak scale. This involves two steps,
first running from the matching scale $M$ down to the top quark mass
and second integrating out the top quark with the anomalous couplings
that are induced by the heavy quark. These two steps are described in
more detail in the next two subsections.

\subsection{Running to $\mu=m_t$}

Given the values of the Wilson coefficients at certain scale, they can
be computed at any other energy (provided no new thresholds are
crossed) by means of the renormalization group equations (RGE). Since
this running is already a loop effect and there are no large logarithms involved, in order to recover the one-loop result in the full theory we just need to include
in the running the effective operators that are generated at tree
level. 
In the model at hand we have \cite{delAguila:2000rc}
\begin{equation}
\mathcal{L}_6^{(0l)} = 
\alpha_{\phi q}^{(1)} \mathcal{O}_{\phi q}^{(1)}
+\alpha_{\phi q}^{(3)} \mathcal{O}_{\phi q}^{(3)}
+\alpha_{u \phi} \mathcal{O}_{u \phi} + h.c.\ ,
\end{equation}
where the operators, following now standard notation, are defined 
\begin{align}%
\mathcal{O}_{\phi q}^{(1)} &= \mathrm{i}\ \phi^\dagger D_\mu \phi\ 
\bar{q} \gamma^\mu q\ , \nonumber \\
\mathcal{O}_{\phi q}^{(3)} &= \mathrm{i}\ \phi^\dagger \sigma^a D_\mu \phi\ 
\bar{q} \gamma^\mu \sigma^a q\ ,\nonumber \\
\mathcal{O}_{u \phi} &= \phi^\dagger \phi\ \bar{q} \tilde{\phi} t\ ,
\end{align}
with coefficients 
\begin{equation}
\alpha_{\phi q}^{(1)} = - \alpha_{\phi q}^{(3)} =
\frac{|\lambda_T|^{2}}{4M^2}, \quad
\alpha_{u \phi} = 2 \lambda_t \alpha_{\phi q}^{(1)}\ .
\end{equation}
The RGE for $\alpha_{1}$ can be found
in \cite{Jenkins:2013wua}, where it is named $\alpha_{\phi D}$, (see
also~\cite{Elias-Miro:2013mua} for further calculations relevant for
the RGE of the SM EFT)  
and reads 
\begin{equation}
16\pi^2 \frac{\mathrm{d}\ \alpha_{1}}{\mathrm{d}\log \mu}
= 8 N_C \lambda_t^2 \alpha_{\phi q}^{(1)} + \ldots\ ,
\end{equation}
where we have only included the contribution proportional to operators
generated at tree level and proportional to the top Yukawa coupling.
In the leading-log approximation we obtain
\begin{align}
\alpha_{1}(m_t) =&~ \alpha_{1}(M) - \frac{N_C \lambda_t^2
  \alpha_{\phi q}^{(1)}(M)}{2 \pi^2}\log\left(\frac{M}{m_t}\right)
\nonumber \\
=&~
\frac{N_C}{32 \pi^2 M^2} %&
\bigg[ \lambda_t^2 |\lambda_T|^{2} -
  |\lambda_T|^{4}
%\nonumber \\ &
%\phantom{
%\frac{N_C}{32 \pi^2 M^2}
%}
-2\lambda_t^2 |\lambda_T|^{2} \log\left(\frac{M^2}{m_t^2}\right)\bigg] .
\end{align}

Using Eq. (\ref{T:from:alphaO1}), we already
recognize the terms proportional to $|\lambda_T|^{4}$ and
$\lambda_t^2 |\lambda_T|^2\log (M^2/m_t^2)$ in
Eq. (\ref{DeltaThat_from_full}). The term proportional to $\lambda_t^2
|\lambda_T|^2$ is not quite right yet but that is just because
we are still missing the second step: integrating out the top quark.

\subsection{Matching at $\mu=m_t$}

In this last step we have to integrate out the top quark. At this
point we have to go to the broken phase of the SM. However, we can
still neglect the bottom mass in our calculation if we want to
reproduce Eq. (\ref{DeltaThat_from_full}). 
Thus, in the new effective theory with the top quark integrated out, 
the relevant fields are massless and the one-loop calculation in the
effective theory side gives vanishing results. 
Thus we only need to perform the computation in the full theory, {\it
  i.e.} in the 
SM. However, due to the terms in $\mathcal{L}_6^{(0l)}$ the top
couplings are modified by terms of order $v^2/M^2$ and these have to
be taken into account.
In particular, the $W_3 t_L t_L$ and $W_1 t_L b_L$
couplings are modified \cite{delAguila:2000rc} (note that
in this reference $v=246$ GeV is used and therefore, there is a
relative factor $\sqrt{2}$ between the corresponding expressions 
there and here):
\begin{align}
g_{W_3 t_L t_L} &= 
g_{W_3 t_L t_L} ^{\mathrm{SM}} [ 1-2 v^2
(\alpha_{\phi q}^{(1)}-\alpha_{\phi q}^{(3)})]
\nonumber \\
&=g_{W_3 t_L t_L} ^{\mathrm{SM}} \left( 1-\frac{|\lambda_T|^2
    v^2}{M^2}\right) , 
\nonumber \\
g_{W_1 t_L b_L} &= 
g_{W_1 t_L b_L} ^{\mathrm{SM}} [ 1+2 v^2
\alpha_{\phi q}^{(3)}]
\nonumber \\
&=g_{W_1 t_L b_L} ^{\mathrm{SM}} \left( 1-\frac{|\lambda_T|^2
    v^2}{2M^2}\right) . 
\label{anomalous:couplings}
\end{align}
Then, the top contribution to $\hat{{\rm T}}$ in the presence of these 
anomalous couplings writes 
(see Ref. \cite{Anastasiou:2009rv}, noting that $\hat{{\rm T}} = {\rm
  T}\ \alpha_{EM}$)  
\begin{align}
\hat{{\rm T}}(m_t^+) =& \frac{N_C}{32 \pi^2 v^2} 
%&
\Bigg \{
\bigg|\frac{g_{W_1 t_L b_L}}{g^{\mathrm{SM}}_{W_1 t_L b_L}}\bigg|^2
\theta_+(m_t,m_b)
\nonumber \\
&-\frac{1}{2}
\bigg[
\bigg|\frac{g_{W_3 t_L t_L}}{g^{\mathrm{SM}}_{W_3 t_L t_L}}\bigg|^2
\theta_+(m_t,m_t) +
\theta_+(m_b,m_b)\bigg]\Bigg\} ,
\label{TSM}
 \end{align}
where $m_t^\pm \equiv \lim_{x\to 0} m_t \pm x$ and
\begin{align}
\theta_+(y_1,y_2) &\equiv y_1^2+y_2^2-\frac{2y_1^2 y_2^2}{y_1^2-y_2^2}
\log\frac{y_1^2}{y_2^2} 
\nonumber \\
&-2\left(y_1^2 \log \frac{y_1^2}{\mu^2} 
+y_2^2 \log \frac{y_2^2}{\mu^2}\right) + \frac{y_1^2+y_2^2}{2} \Delta\ .
\nonumber 
\end{align}
We temporarily keep the bottom mass to regulate IR divergencies. 
Using the explicit values for the anomalous couplings in
Eq. (\ref{anomalous:couplings}) and taking the bottom mass to zero 
we get
\begin{align}
\hat{{\rm T}}(m_t^+)&=
\frac{N_C}{32 \pi^2 v^2} 
\bigg \{
\bigg(1-\frac{|\lambda_T|^2v^2}{M^2}\bigg)\theta_+(m_t,0)
\nonumber \\
&\qquad-\frac{1}{2}
\bigg(1-\frac{2|\lambda_T|^2v^2}{M^2}\bigg)\theta_+(m_t,m_t) 
\bigg\}
\nonumber \\
&=
\frac{N_C}{32 \pi^2} \lambda_t^2\left(1-\frac{|\lambda_T|^2
v^2}{M^2} \right) = \hat{\rm T}_{\mathrm{SM}} + \Delta \hat{\rm T}(m_t^+)\ ,
 \end{align}
where we have explicitly split the result into the SM contribution, 
$\hat{{\rm T}}_{\rm SM}$, and a correction, 
$\Delta\hat{{\rm T}}(m_t^+)$, which is 
proportional to $|\lambda_T|^2$. 
We have used that 
\[
\theta_+(m,0)=m^2[1-2\log(m^2/\mu^2)+\Delta/2],
\] 
and
\[
\theta_+(m,m) = m^2[-4 \log(m^2/\mu^2)+\Delta],
\] 
and set $\mu=m_t$ to eliminate
the logarithms. 
We have only kept terms up to $\mathcal{O}(v^2/M^2)$ and used the $\overline{\rm MS}$
renormalization scheme to remove the divergent term left, proportional
to $\Delta$ and $|\lambda_T|^2$, with the corresponding counterterm.

Since the contribution below $m_t$ vanishes if we neglect the bottom
mass, we have ended the calculation. Putting all pieces
together: 
\begin{align}
\Delta\hat{{\rm T}}(m_t^-)
&=- v^2 \alpha_{\phi D}(m_t)+\Delta\hat{{\rm T}}(m_t^+) 
\nonumber \\
&=
\frac{N_C }{32 \pi^2}\frac{v^2}{M^2} 
\left[ 
 |\lambda_T|^4
+2\lambda_t^2 |\lambda_T|^2
\left(\log\frac{M^2}{m_t^2}
-\frac{1}{2}\right)
\right]
%\nonumber \\&
-\frac{N_C }{32 \pi^2}\frac{v^2}{M^2} 
\lambda_t^2 |\lambda_T|^2
\nonumber \\
&=
\frac{N_C }{32 \pi^2}\frac{v^2}{M^2} 
\left[ 
 |\lambda_T|^4
+2\lambda_t^2 |\lambda_T|^2
\left(\log\frac{M^2}{m_t^2}
-1\right)
\right],
\end{align}
which is exactly Eq. (\ref{DeltaThat_from_full}).

\end{document}